\documentclass[journal]{IEEEtran}

\usepackage{cite}

\ifCLASSINFOpdf
   \usepackage[pdftex]{graphicx}
   \graphicspath{{../pdf/}{../jpeg/}}
   \DeclareGraphicsExtensions{.pdf,.jpeg,.png}
\else
   \usepackage[dvips]{graphicx}
   \graphicspath{{../eps/}}
   \DeclareGraphicsExtensions{.eps}
\fi

\usepackage{subfig}

\begin{document}
\title{Channel Static Antennas}

\author{Gerald Artner,~\IEEEmembership{Member,~IEEE}
\thanks{Gerald Artner was with the Institute of Telecommunication, Technische Universit\"at Wien, Gu\ss hausstra\ss e 25, 1040 Vienna, Austria, e-mail: gerald.artner@nt.tuwien.ac.at, website: geraldartner.com.}
\thanks{Manuscript received March 14, 2018; revised August 26, 2015.}}

\markboth{Journal of \LaTeX\ Class Files,~Vol.~14, No.~8, March~2018}%
{Shell \MakeLowercase{\textit{et al.}}: Bare Demo of IEEEtran.cls for IEEE Journals}

\maketitle

\begin{abstract}
The possibility to keep wireless communication channels static is investigated.
When an antenna is moved away from its position, this will in general cause the channel to change.
Considerations suggest that wireless communication channels can be kept static by performing a counter-movement of the antenna to keep it in its original position relative to outside observers.
Feasibility is shown for a platform moving in straight motion over a finite distance.
The channel is kept static by countering the platform's movement with physical movement of the antenna in the opposite direction.
The experiment is conducted with a quarter-wavelength monopole antenna in the gigahertz range.
\end{abstract}


\section{Introduction}

\IEEEPARstart{S}{tatic} communication channels are a prerequisite for modern wireless transmission schemes, which rely on the channel being static during the time it takes for a transmission, or for the time period between channel estimation and transmission. 
However, in many applications antennas are mounted on moving platforms.
Prominent examples are user equipments in mobile communications and automotive antennas in vehicular communications.
Even if the environment is static otherwise, platform movement generally changes the amplitude, phase and frequency of the transmitted and/or received electromagnetic waves.
In turn, this changes the communication channel and it becomes time-variant, even if the initial channel without the platform's movement would not be time-variant.
Modern communication schemes that use the channels between multiple antennas suffer severely from changing channels \cite{Artner2013}.
Especially, vehicular channels become time-variant due to antenna movement \cite{Mecklenbrauker2011ProcIEEE}.

Attempts to keep communication channels static have been undertaken previously.
Several concepts aim to keep wireless channels constant between estimation and transmission.
Examples are the Predictor Antenna \cite{Sternad2012,Bjorsell2017} and the Separate Receive and Training Antenna \cite{phanhuy2013} concepts.
Both concepts introduce a single antenna which is used at the time when the channel is estimated and several successive antennas that are used to transmit data.
The channel becomes similar to the estimated channel at discrete times when a successive antenna reaches the position where the first antenna was at the time of channel estimation.
This requires that the antenna switching apparatus is aware of the transmission scheme in order to divert the channel estimation sequence or symbol to the first antenna.
Previous authors also propose to alter the transmission scheme such that payload data is only transmitted at times when a successive antenna reaches the position of the first one.

A conformal antenna array for the conical tips of projectiles is used in \cite{Jaeck2017} to compensate the projectile's rotation around its axis of revolution.

Considerations in this work suggest that wireless communication channels can be kept static by performing a counter-movement of the antenna.
The channel static antenna concept is evaluated experimentally for finite linear motion.
Measurements are performed in an anechoic chamber and in an office environment.

\section{Gedankenexperiment}

The following considerations suggest that communication channels can be kept static by performing a counter-movement of the antenna on the platform.

\begin{figure}[!t]
\centering
\subfloat[]{\includegraphics[width=0.13\textwidth]{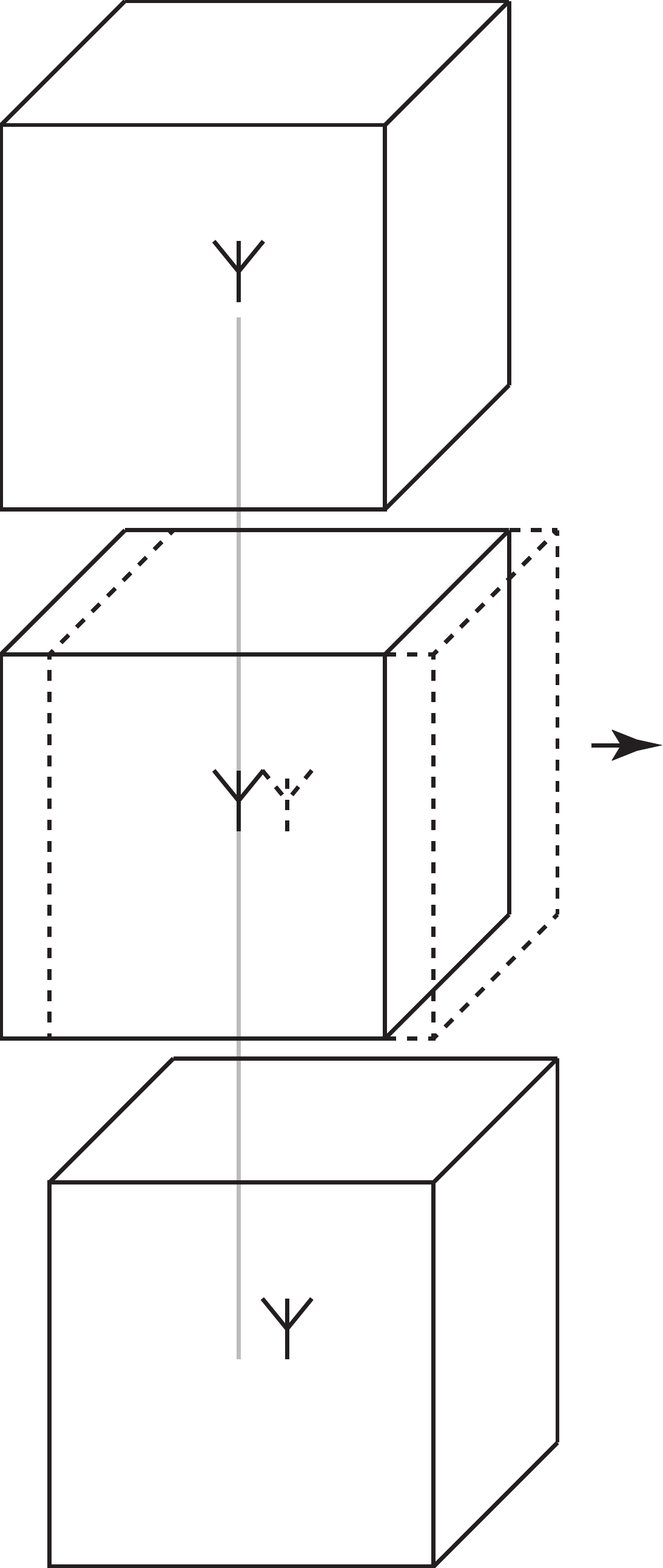}
\label{fig_volume_regular}}
\hfil
\subfloat[]{\includegraphics[width=0.13\textwidth]{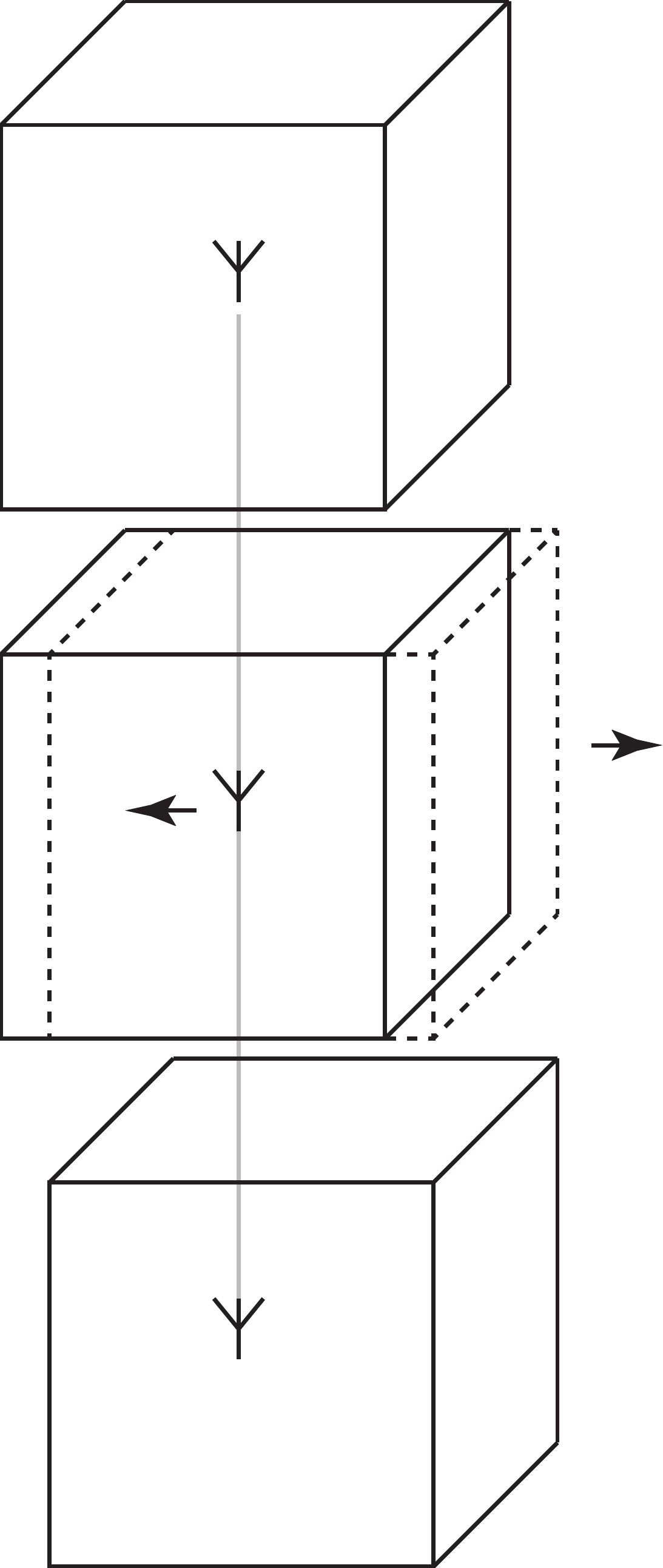}
\label{fig_volume_moved}}
\caption{a) First gedankenexperiment. \emph{Top:} An antenna is mounted in a volume. \emph{Center:} The volume is then moved. \emph{Bottom:} This movement would generally change the channel to a second antenna (not depicted) at an arbitrary position.\\
b) Second gedankenexperiment. \emph{Top:} An antenna is mounted in a volume. \emph{Center:} The volume is then moved, but the antenna counters this by moving in the opposite direction. \emph{Bottom:} The channel stays static because nothing has changed; except that an imaginary volume was moved.}
\label{fig_volume}
\end{figure}

Consider the following gedankenexperiment sketched in Fig.~\ref{fig_volume_regular}.
A volume of arbitrary shape and size is defined in empty space.
An antenna is mounted inside the volume, but the volume stays empty otherwise.
The volume is then moved.
Because the antenna is mounted to the volume it moves with it.
For a second antenna, it is expected that the amplitude, phase and frequency of the received signal change due to the movement.
The wireless communication channel between these two antennas changes.

Now consider the related gedankenexperiment sketched in Fig.~\ref{fig_volume_moved}.
Again a volume of arbitrary shape and size is defined in empty space, an antenna is mounted inside the volume and the volume stays empty otherwise.
The volume is again moved, but the antenna counters the movement by simultaneously moving into the exact opposite direction with the exact same speed.
A second antenna at an arbitrary position would experience no changes in amplitude, phase or frequency, because the first antenna stayed at its original position.
Therefore, the communication channel between these two antennas stays the same --- it is kept static.

\begin{figure}[!t]
\centering
\includegraphics[width=0.30\textwidth]{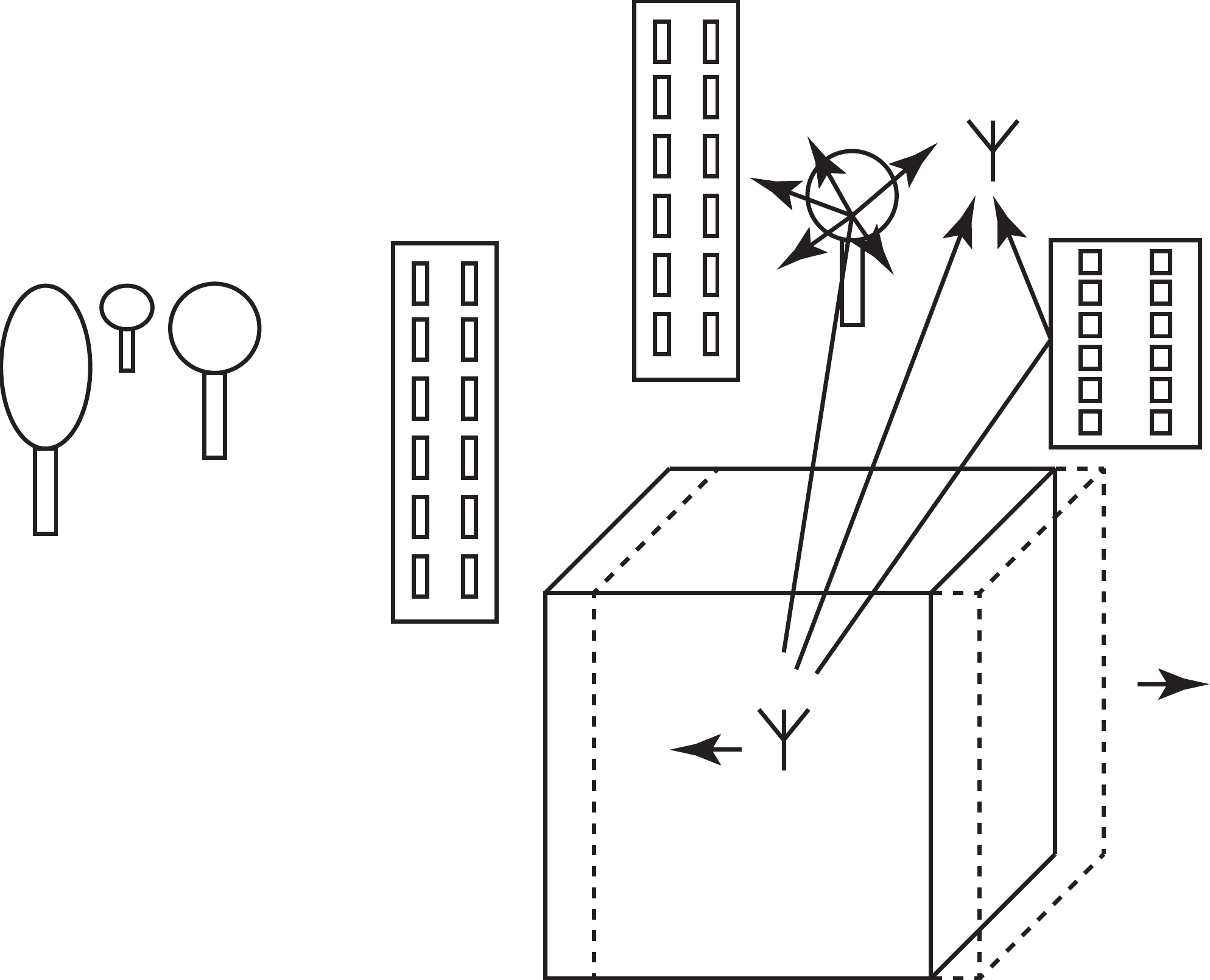}
\caption{Even with the addition of complex objects, the channel stays static, as long as the objects don't move or change relevant properties.}
\label{fig_volume_clutter}
\end{figure}

The gedankenexperiment is now evolved by \textit{seemingly} adding complexity in Fig.~\ref{fig_volume_clutter}.
Objects are added that reflect electromagnetic waves, that scatter them, absorb, refract, amplify and so on.
The channel becomes more complex in this environment, but the antenna's ability to keep the channel static is unimpeded.
The channel stays static as long as the antenna can counter the volume's movement and as long as the objects don't move or change their relevant properties.

\subsection{Practical Considerations}

\begin{figure}[!t]
\centering
\includegraphics[width=0.40\textwidth]{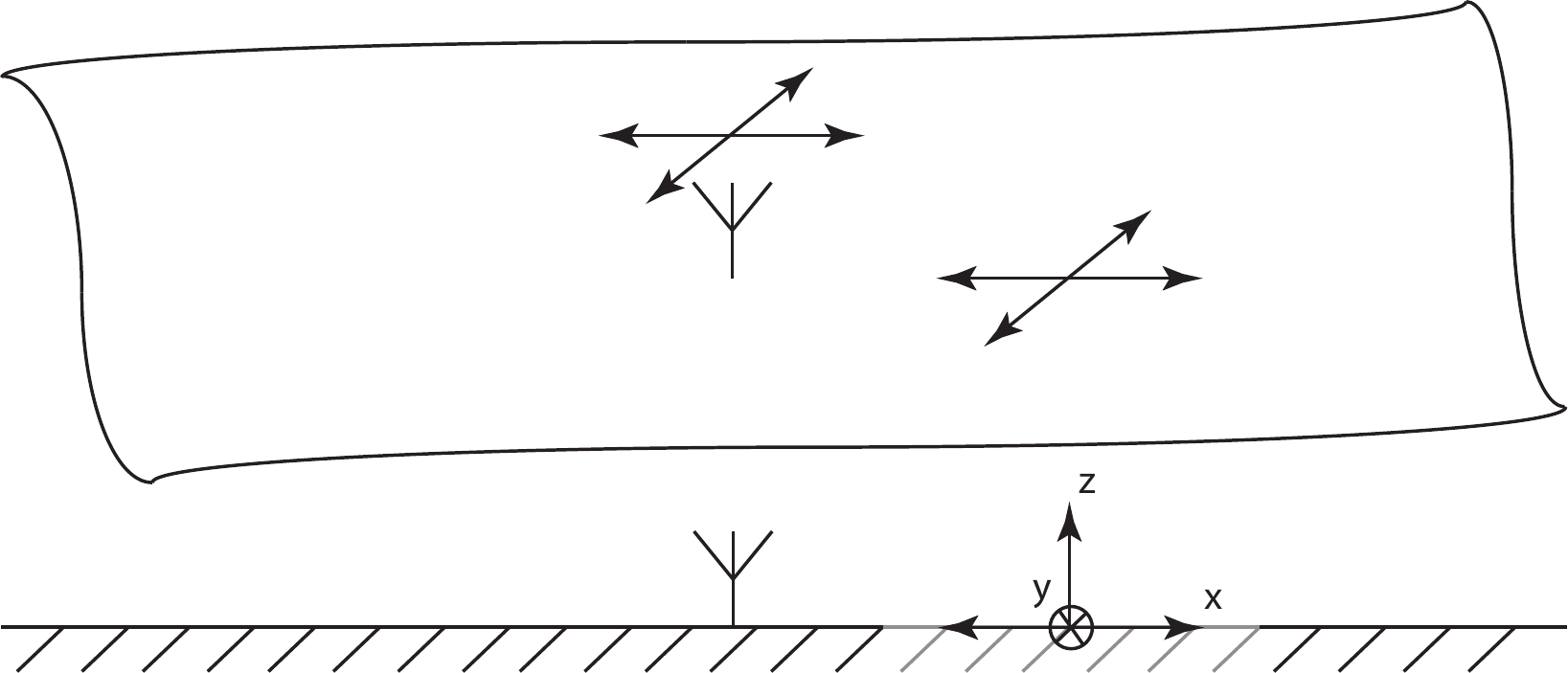}
\caption{An antenna is placed on a perfect conductive sheet of infinite size. It can keep the channel static under sheet movement in x-y plane by performing a counter-movement on its surface.}
\label{fig_groundplane}
\end{figure}

An antenna inside an imaginary volume in infinite empty space might be of interest in space applications.
In most other applications it is desired to mount the antenna on a moving platform.
The gedankenexperiment is therefore modified towards applications.
Consider an antenna that is mounted on a perfect conducting sheet of infinite size (e.g. in x-y plane, see Fig.~\ref{fig_groundplane}).
Let the sheet now move in x-y plane.
Again, if the antenna counters the movement of the plane by moving in the opposite direction, then it stays in the same place from the view of an outside antenna and the channel remains unchanged.
Objects can be added to the space above the ground plane analogous to Fig.~\ref{fig_volume_clutter} with the same argument.
Although still theoretical, this consideration leads towards more practical systems.
The space underneath the sheet is entirely irrelevant to the channel, because electromagnetic waves can not penetrate it.
From a design viewpoint, the space below the sheet can house arbitrary components.
Antennas mounted on large ground planes with finite conductivity are in widespread commercial use.
Examples are monopole antennas mounted on vehicle roofs and inverted-F antennas mounted on mobile phones.

A finite size of the moving platform limits the distance that an antenna can be moved.
Generally trajectories will be longer than the size of the object - this limits the time that the channel can be kept static with this technique.
As a coincidence, drag considerations of moving platforms in gases and fluids result in airplanes, cars, trains, submarines and rockets that are designed to be long in the direction of their movement.
Additionally, many objects exclusively move in the direction of their hulls, e.g. in normal operation an antenna on the car roof will not move up or down and a rocket will not move sideways.
Most devices already contain apparatuses to measure acceleration, velocity and movement direction.
The trend towards higher frequency bands results in antennas that are small compared to the platforms that they are mounted on, and it means that even small areas like a car roof becomes tens or hundreds of wavelengths long.
Even user equipment for cellular networks is large enough to compensate small movements by the user.
The paradigm of automation even results in planned trajectories, which makes the counter-movements of antennas predictable.
Although antenna movement is trivial, physically moving antennas might be undesirable in practice and alternative solutions should be developed.

As exemplary calculations, consider antennas mounted on a car and a cellular phone.
Assuming the length of a typical car as $4$\,m \cite{automobiledimensions} and considering a speed limit of $150$\,km/h or $42$\,m/s, it would be then possible to keep channels constant for $95$\,ms.
Assuming only $1$\,cm for antenna movement on a smart phone at a speed of $20$\,km/h, it would be possible to keep channels constant for $1.6$\,ms.
These times are long e.g. when compared to the $0.5$\,ms slot duration of LTE.

\section{Experimental Evaluation of Channel Static Antennas Under Finite, One-Dimensional Platform Movement}

The gedankenexperiment is altered for the first experimental evaluation.
Without loss of generality, the platform movement is limited to a straight line.
The infinitely long line is further replaced by a platform of finite length.
The movement of this platform is limited to finite lengths that are shorter than the platform, such that the antenna can counter the platform's movement by physically moving into the opposite direction (without falling off the platform).

\begin{figure}[!t]
\centering
\includegraphics[width=0.40\textwidth]{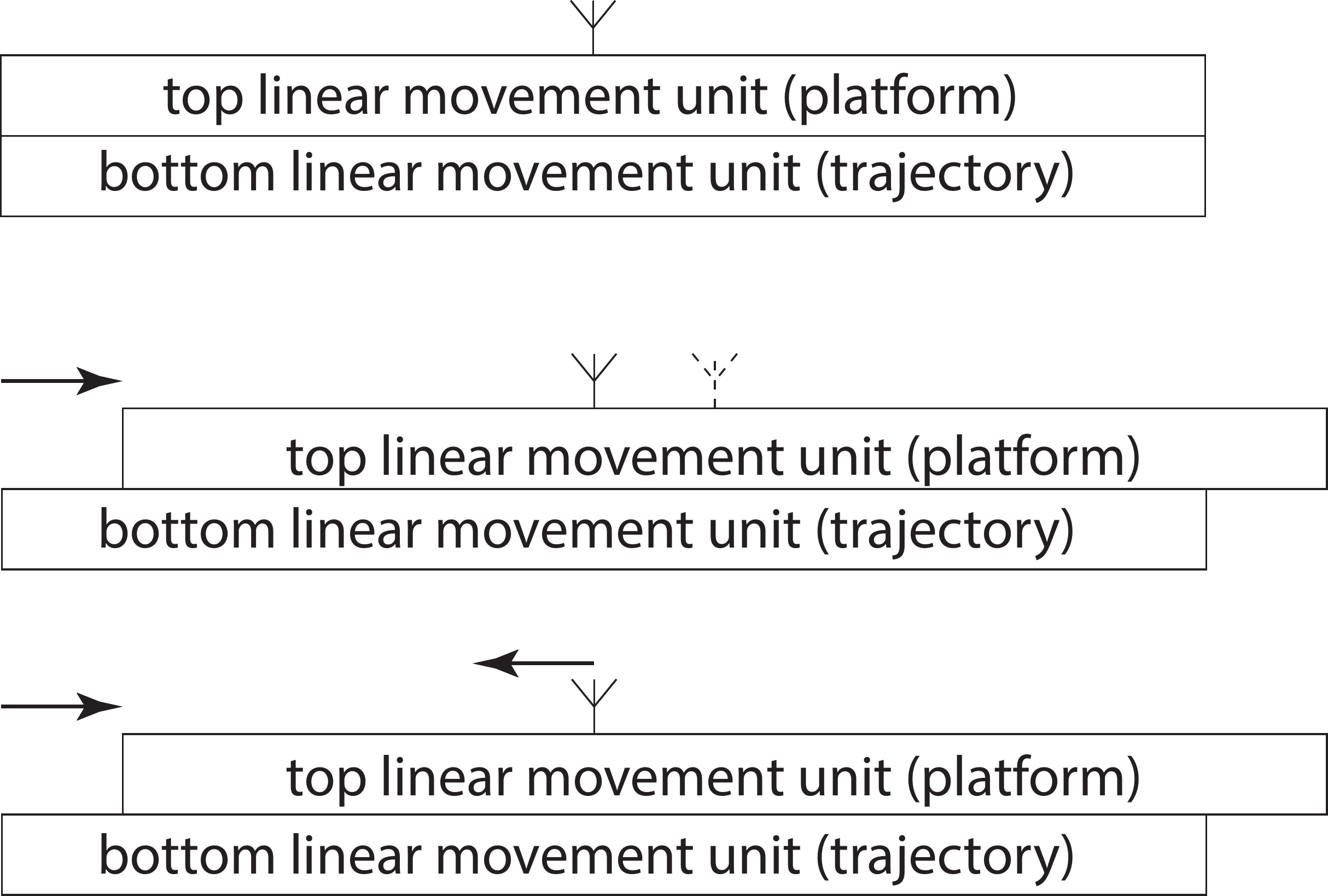}
\caption{A sketch of the experiment. An antenna is mounted on two linear movement units (top). The top unit acts as a moving platform that the antenna is mounted on (middle). The top unit moves the antenna in the opposite direction, such that it stays in its original position (bottom).}
\label{fig_sketch}
\end{figure}

The experiment is conducted on the Vienna MIMO Testbed \cite{Caban2006}.
The platform is realized as a linear movement unit.
The antenna is mounted on this platform and a second linear movement unit is mounted underneath it.
The bottom unit moves the platform and the top unit moves the antenna to counter the platform's movement.
The experiment is sketched in Fig.~\ref{fig_sketch}.

The antenna is built as a quarter-wavelength monopole antenna.
The perfect conductive ground plane of infinite size is approximated by an aluminum sheet with a diameter of $180$\,mm ($\approx 1.5$\,$\lambda$).
The antenna is elevated by a $26$\,cm long column such that a coaxial cable can be connected to the subminiature version-A (SMA) connector on the bottom.
A second quarter-wavelength monopole antenna is placed in the same room at a distance of $\approx 2$\,m.
Both antennas are connected to a vector network analyzer (VNA) with coaxial cables.
Narrowband channel measurements are performed in the industrial, scientific and medical (ISM) band at $2.45$\,GHz.
The measurements are automated by controlling the platform movement, the antenna movement and the VNA with a laptop.
The platform is moved in steps, because the VNA measurements take some time.
In a step the platform is moved $0.02$\,$\lambda$, then the setup waits for $0.2$\,s to wear off vibrations from acceleration/deceleration, then the VNA measurement is performed.
Therefore, Doppler shift \cite{Doppler1842} is not considered.

\subsection{Experiment in Anechoic Chamber}

\begin{figure}[!t]
\centering
\subfloat[]{\includegraphics[width=0.49\linewidth]{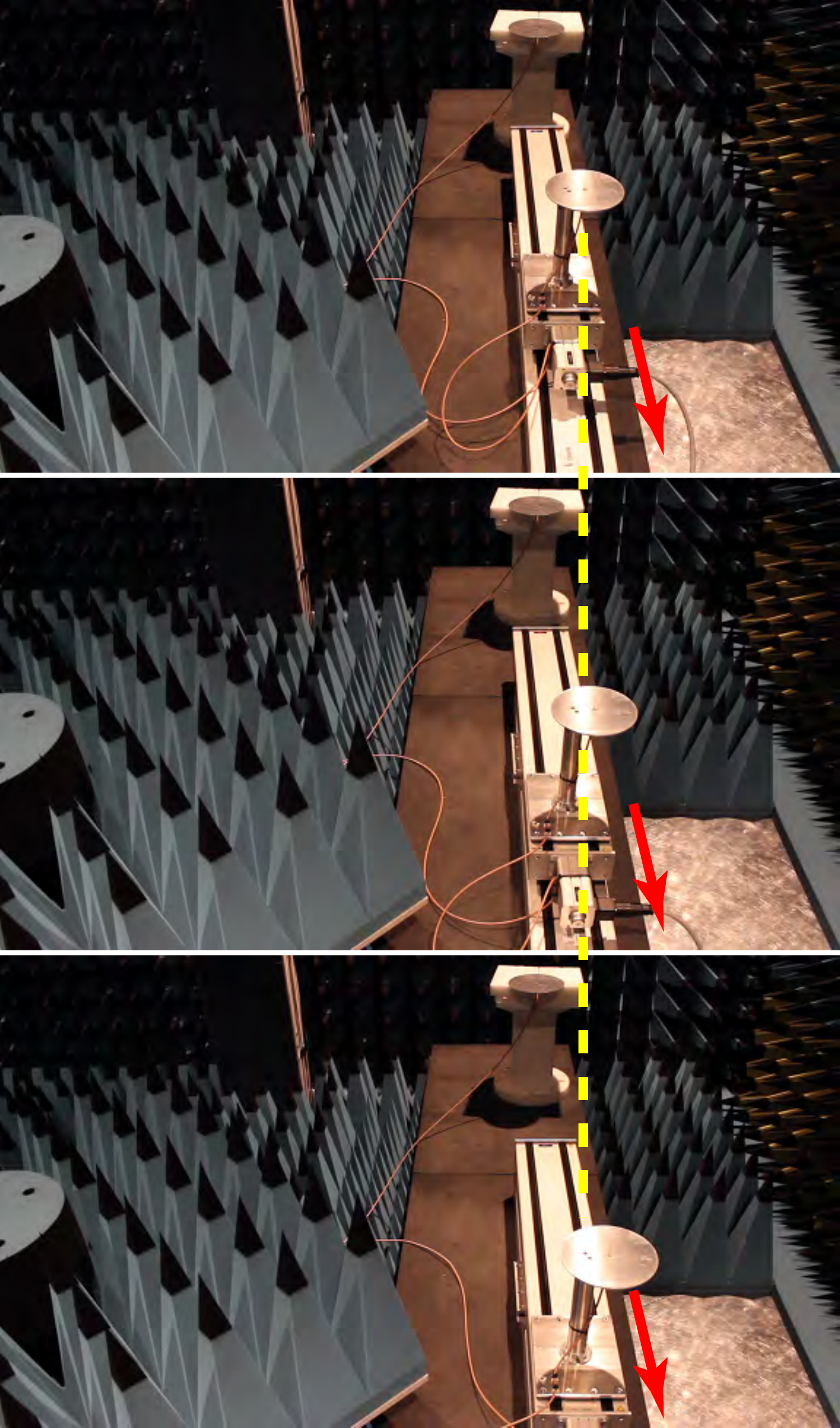}}
\hfil
\subfloat[]{\includegraphics[width=0.49\linewidth]{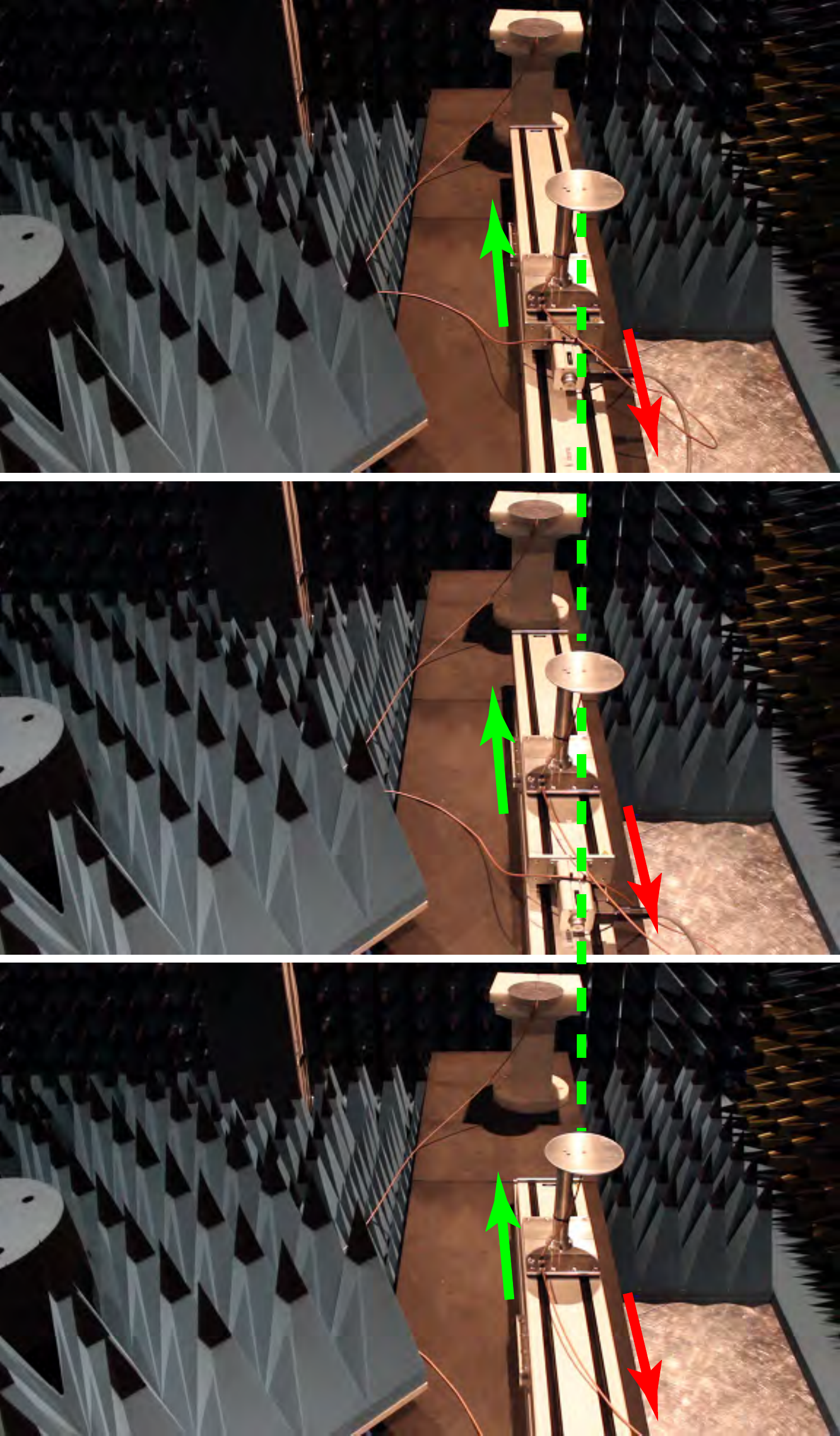}}
\caption{A platform moves towards the front in an anechoic environment. a) The antenna is fixed on the platform and moves with it. b) The antenna counters the platform movement to keep its position relative to the surroundings static.}
\label{fig_chamber_photo}
\end{figure}

The anechoic chamber is a simple environment in terms of wireless channels.
Multi-path propagation is heavily attenuated and as a result there are no delayed taps, no angle-of-arrival, and no small-scale fading.
Amplitude is expected to vary little over antenna position.
Three measurements are performed.
First, the platform is moved over a distance of $6$\,$\lambda$ ($\approx 73$\,cm).
The antenna is fixed to the platform and moves with it.
Second, the platform is moved over $6$\,$\lambda$ and the antenna counters the platform's movement by performing a counter-movement in the opposite direction.
Third, both the antenna and the platform remain still in the initial position.
The third measurement acts as a reference to quantify the channel changes caused by the environment, which are not compensated in the presented work.
Fig.~\ref{fig_chamber_photo} shows photographs of the experiment.

\begin{figure}[!t]
\centering
\subfloat[]{\includegraphics[width=0.40\textwidth]{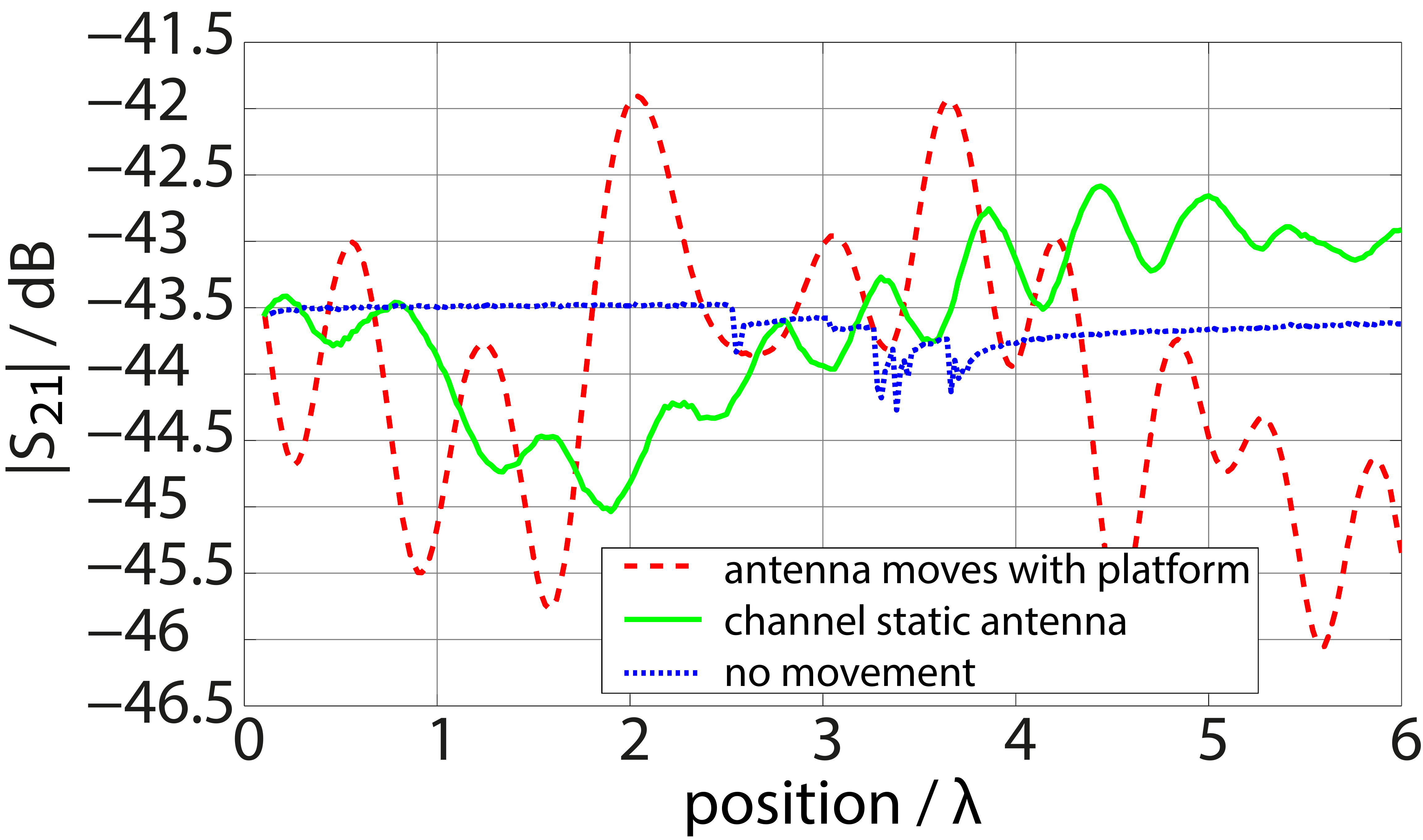}\label{fig_measurements_chamber_amplitude}}
\\
\subfloat[]{\includegraphics[width=0.40\textwidth]{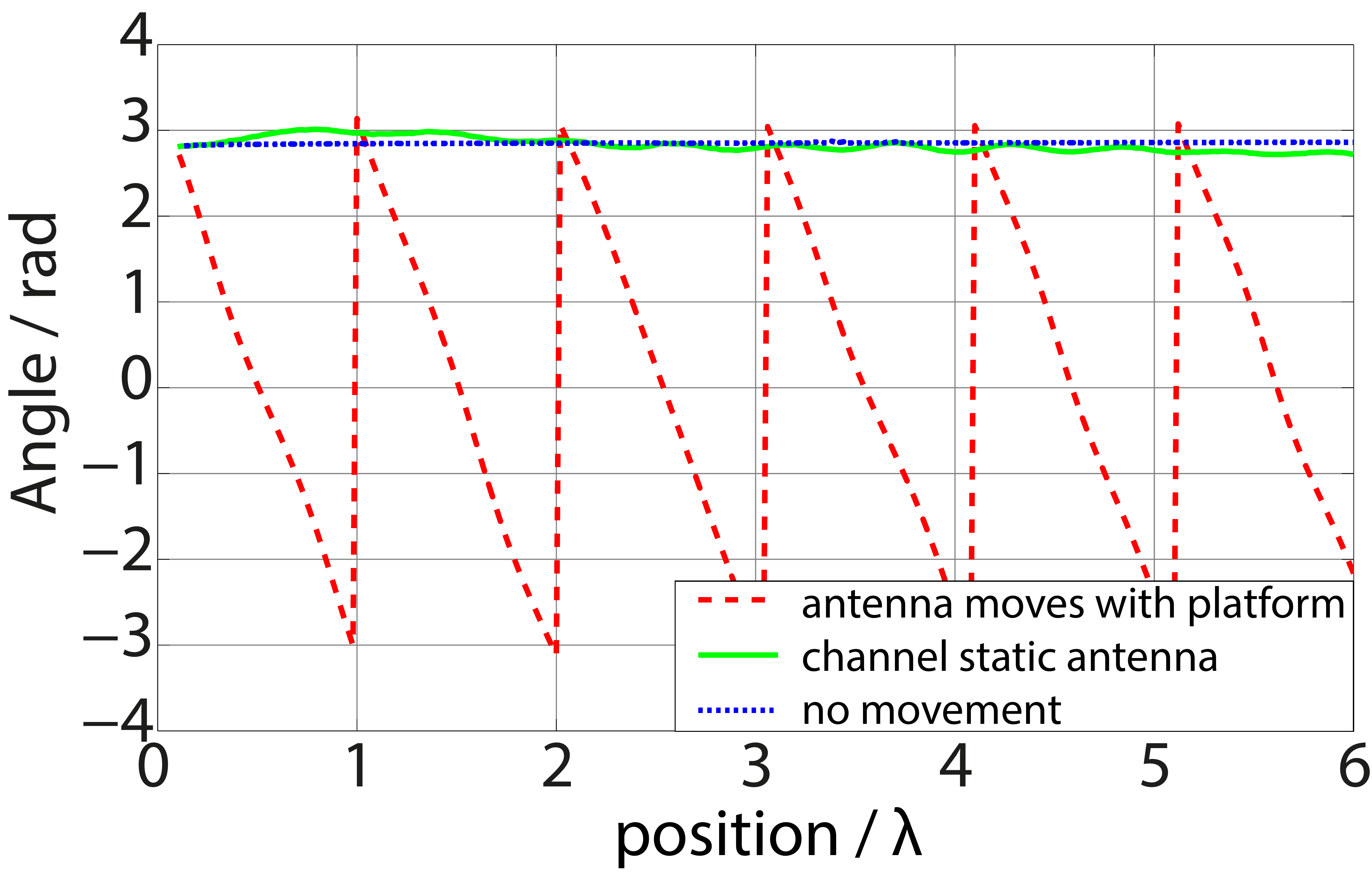}\label{fig_measurements_chamber_phase}}
\caption{Measurement results of the channel static antenna inside the anechoic chamber: a) amplitude and b) phase. The measurement without movement is shown as reference. The antenna stands in the initial position for a time period similar to the other two measurements.}
\label{fig_measurements_chamber}
\end{figure}

The measured scattering-parameters (S-parameters) of the channel inside the anechoic chamber are shown in Fig.~\ref{fig_measurements_chamber}.
The ``ordinary'' moving antenna and the channel static antenna are plotted as a function of the length in wavelengths that they moved away from their initial positions.
A measurement without movement is shown as a reference.
This measurement is plotted over time for a similar duration as the two other measurements.

Fig.~\ref{fig_measurements_chamber_amplitude} shows the absolute value of the measured scattering parameters. 
As expected in an anechoic environment, the power variations are quite small.
There is small residual wave interference from reflections which causes changes in the absolute values of the scattering parameters of $\pm 2$\,dB.
The overall variation and the ripple is smaller with the channel static antenna than with the regular moving antenna, but there is significantly more variation in amplitude than in the reference measurement without movement.
This residual change with the channel static antenna is attributed to the technical solution that performs the counter-movement of the antenna.
In this experiment this is a linear movement unit that protrudes underneath the antenna and therefore influences the channel.
It is not clear what caused the changes in the measurement without antenna movement.
Similar measurements without antenna movement showed no such changes.

Fig.~\ref{fig_measurements_chamber_phase} shows the phase between the two antennas.
The phase is wrapped over $2\pi$ in the plot such that the results without movement and with the channel static antenna are still visible.
As expected in an anechoic environment, the channel changes are primarily revealed in the phase difference between the two antennas.
The channel static antenna keeps the phase almost perfectly static.
It does so over a long distance of platform movement ($6$\,$\lambda$).

\subsection{Experiment in Office Environment}

\begin{figure}[!t]
\centering
\includegraphics[width=0.40\textwidth]{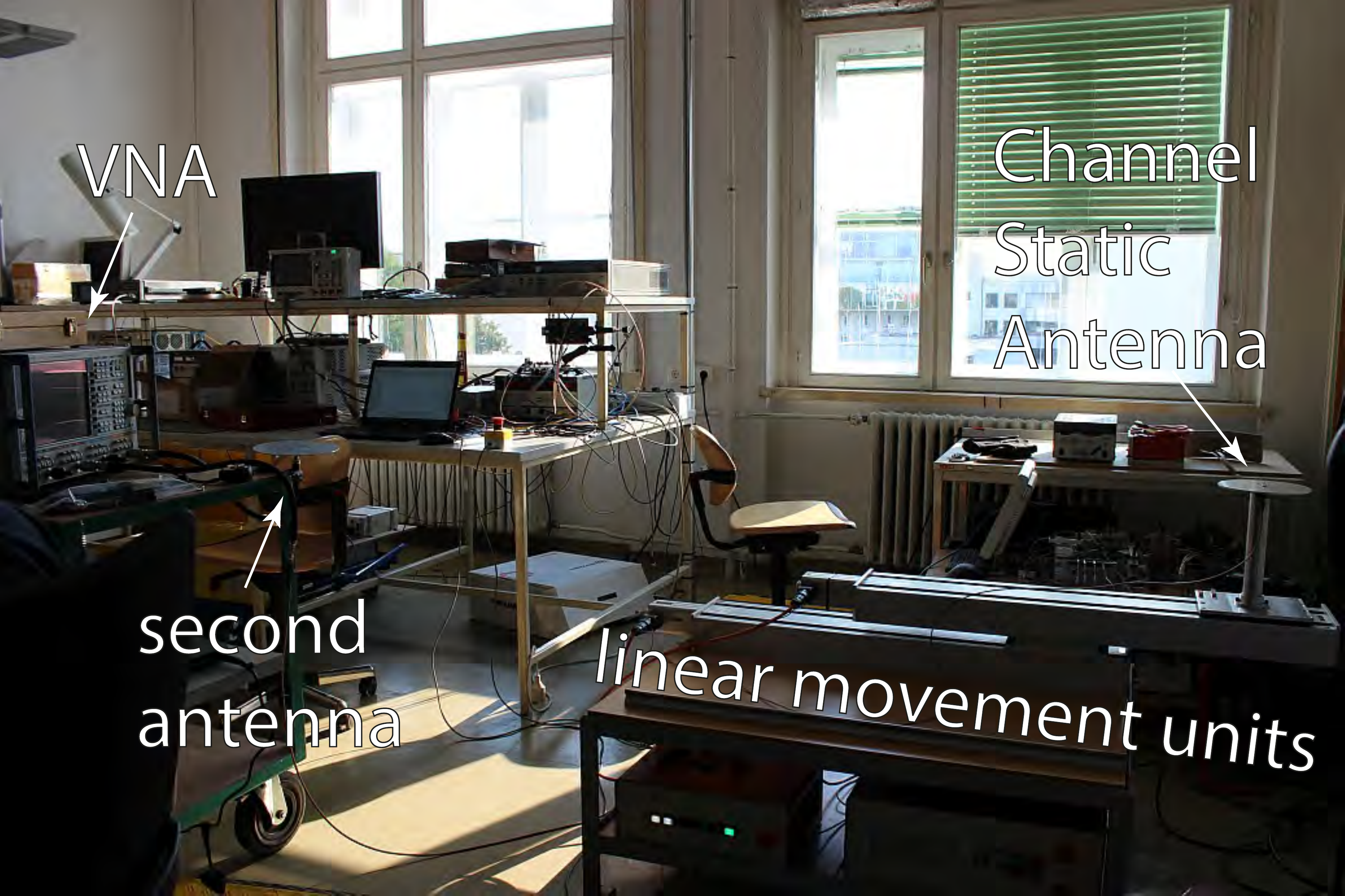}
\caption{Photograph of the measurement setup and the surrounding office environment. Notice the complex shapes of the objects and the variety of different materials.}
\label{fig_raum}
\end{figure}

The experiment is repeated in a laboratory/office environment.
This environment contains a large number of different materials that are arranged in complex shapes (see Fig.~\ref{fig_raum}).
No people were present in the room during the measurements.

\begin{figure}[!t]
\centering
\subfloat[]{\includegraphics[width=0.49\linewidth]{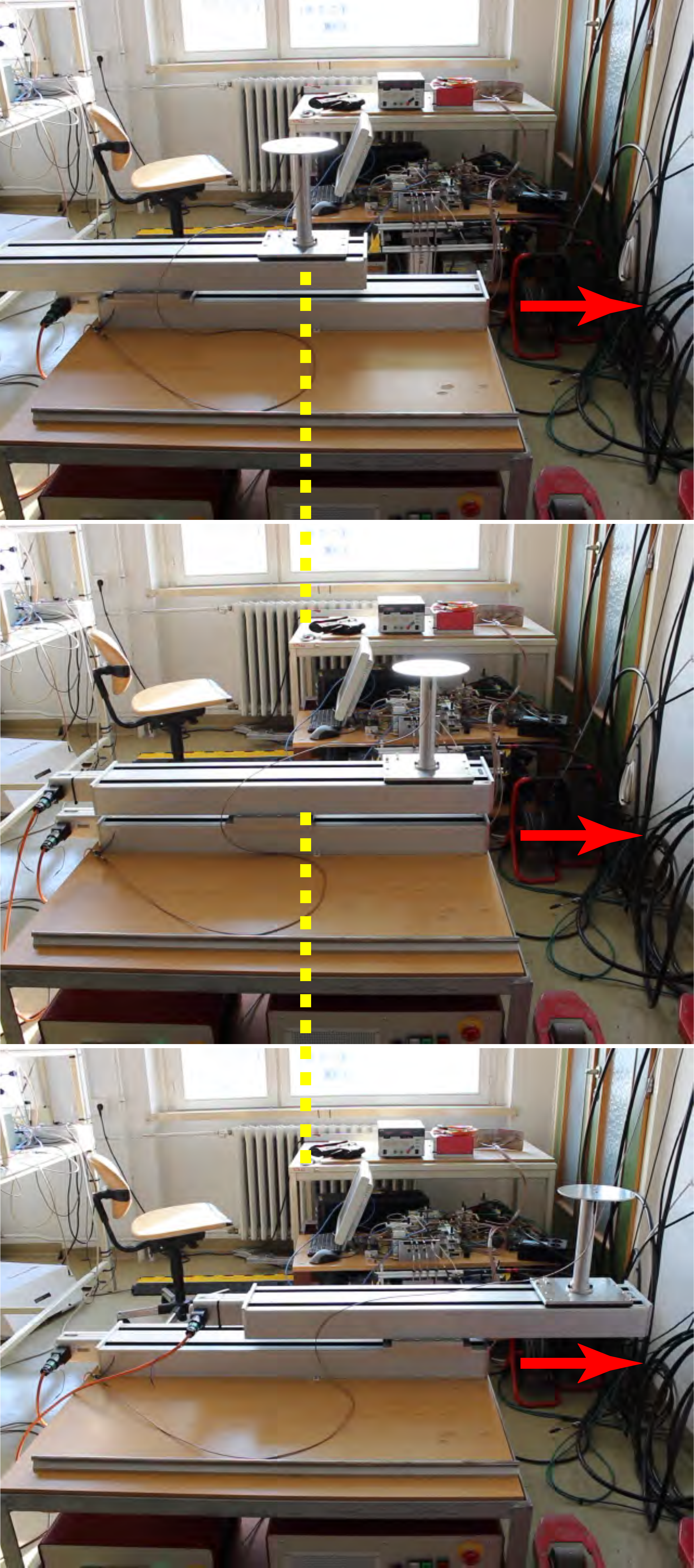}}
\hfil
\subfloat[]{\includegraphics[width=0.49\linewidth]{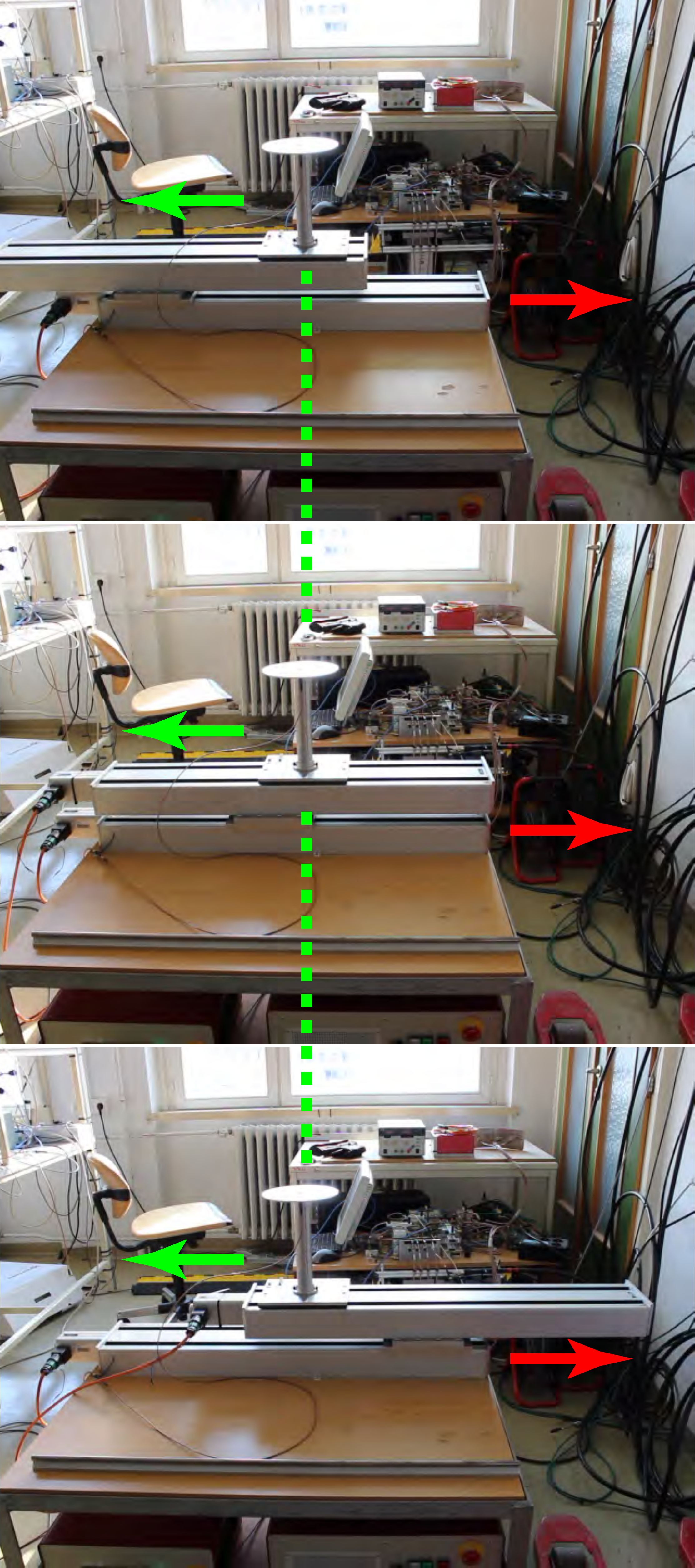}}
\caption{A platform moves from left to right in an office environment. a) The antenna is fixed on the platform and moves with it. b) The antenna counters the platform movement to keep its position relative to the surroundings static.}
\label{fig_measurements}
\end{figure}

Again, three measurements are performed.
First, the antenna is fixed to the platform and moves with it.
Second, the antenna is mounted to the platform and it counters the platform's movement with a counter-movement in the opposite direction.
Third, both the antenna and the platform remain still to quantify environmental influences.
Fig.~\ref{fig_measurements} shows the setup at selected times during the measurement.

\begin{figure}[!t]
\centering
\subfloat[]{\includegraphics[width=0.40\textwidth]{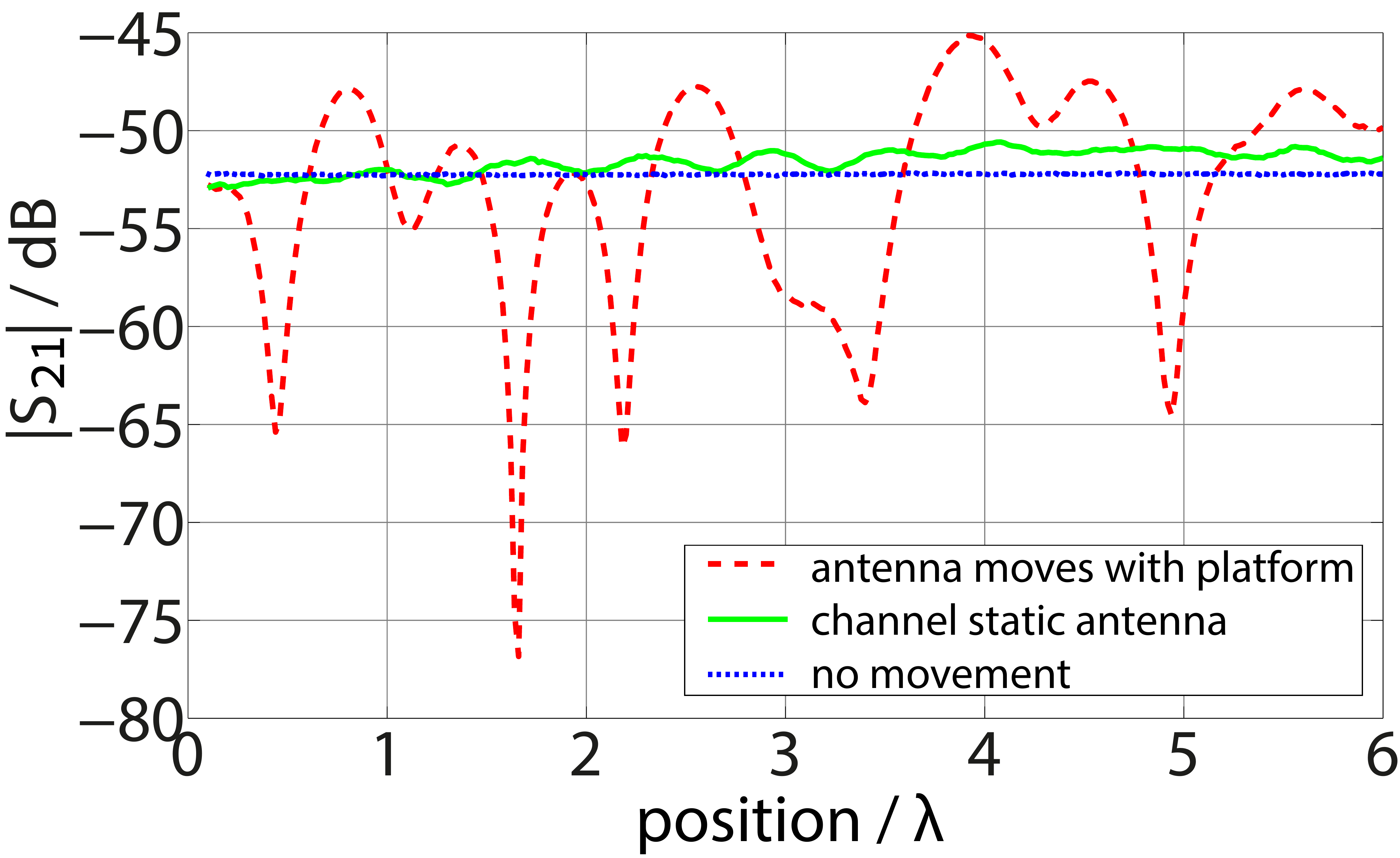}\label{fig_measurements_office_amplitude}}
\\
\subfloat[]{\includegraphics[width=0.40\textwidth]{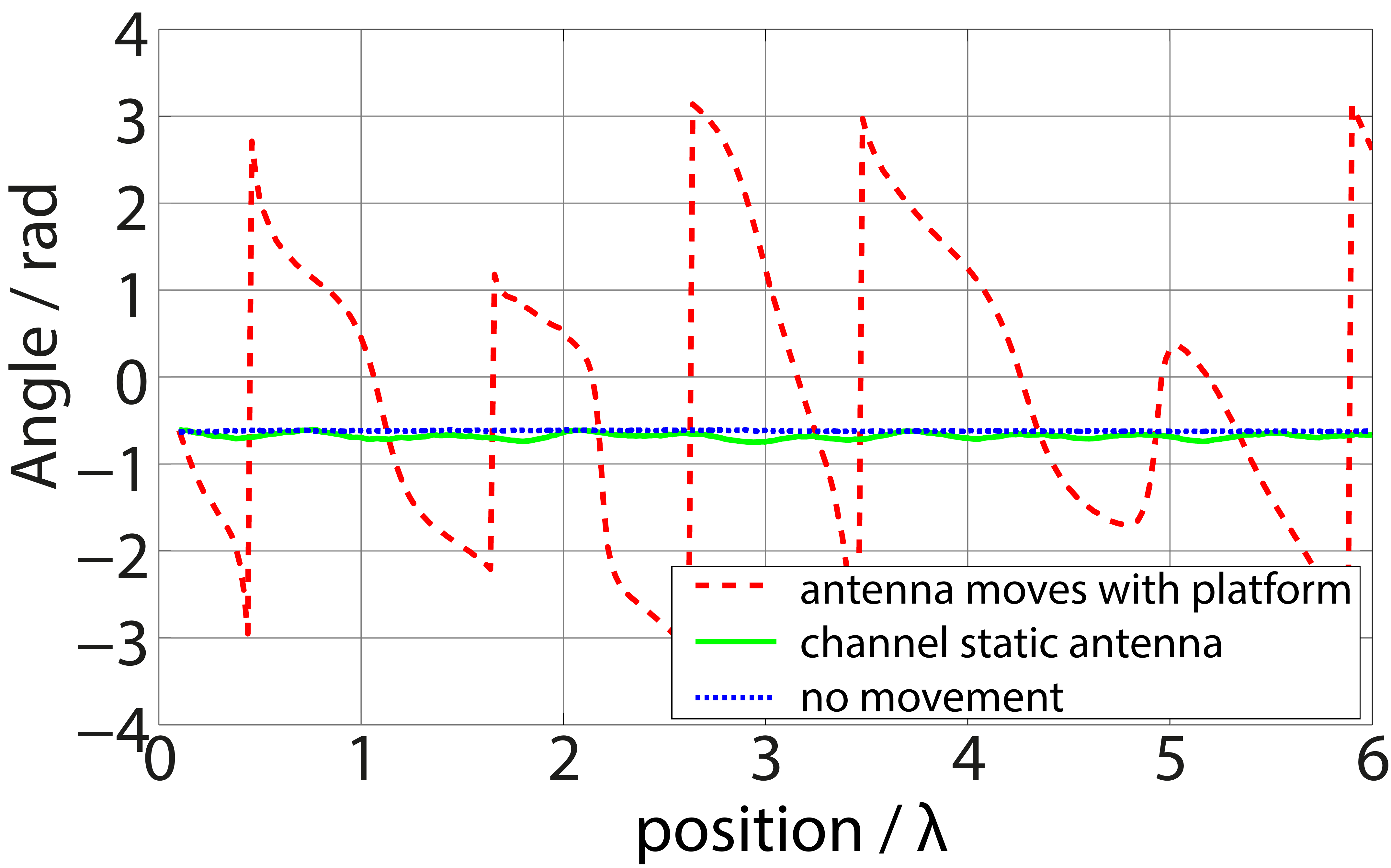}\label{fig_measurements_office_phase}}
\caption{Measurement results of the channel static antenna inside the office: a) amplitude and b) phase. The measurement without movement is shown as reference. The antenna stands in the initial position for a time period similar to the other two measurements.}
\label{fig_measurements_office}
\end{figure}

The absolute value of the measured S-parameters between the two antennas is shown in Fig.~\ref{fig_measurements_office_amplitude}.
The measured phase of the S-parameters is shown in Fig.~\ref{fig_measurements_office_phase} and it is again wrapped for convenient depiction.
Even thought the office environment has a complex geometry and a variety of materials, the communication channel is not time-variant on its own.
It stays practically the same without antenna movement.
When the antenna is fixed to the platform and moves with it, the measured channel is characterized by the deep fading notches that are typically observed in office environments.
The changing channel is a direct result of the antenna's movement through the multipath environment with constructive and destructive interference.
The channel changes when the antenna platform is moved through the small scale fading environment, which is then observed as fast fading by the moved antenna.

Now consider the measured curve where the platform movement is compensated by moving the antenna in the opposite direction of the platform's movement.
Small scale fading is no longer present and the channel is approximately static, because the antenna no longer moves through the multipath interference pattern.
Without movement the channel changes only by $0.16$\,dB, while it changes by $31.69$\,dB with the regular antenna and $2.35$\,dB with the channel static antenna (all values min to max).
The channel static antenna does not keep the channel completely static in this scenario.
The movement unit (platform) of the channel static antenna still moves through the electromagnetic field, and by doing so distorts it.
After all, the antenna is not perfectly shielded from the linear movement unit by an infinite ground plane.
The phase is kept almost completely static by the channel static antenna.
In the measured scenario, the channel static antenna keeps the phase within $8.39^\circ$ over a platform movement of $6$\,$\lambda$.

\section{Conclusion}

The above considerations suggest that electromagnetic waves between a transmitter and a receiver moving relative to each other can be kept static in amplitude, phase and frequency with proper antenna design.
This suggests that Doppler Shift \cite{Doppler1842} is not an intrinsic property of moving objects, but that it can be compensated.
The channel is kept completely static in cases where changes were only caused by antenna movement.
This should be approximately valid in many environments.
The presented considerations do not compensate channel changes that are caused by other objects.

The channel static antenna concept is proven to be feasible.
Experimental results are presented and discussed for an anechoic environment and an office environment.

The channels in the investigated scenarios are intrinsically static.
Even in the office, wave propagation is complex, but the channel remains largely static without the presence of moving scatterers (e.g. people).
Changes to these channels are caused by the antenna's own movement.
It is state-of-the-art to accept this.
The presented measurements proof that wireless communication channels can be kept static by compensating the movement of the platform that they are mounted on.

State-of-the-art transmission schemes accept that communication channels change.
When the channel gets worse, then the received amplitude drops below the initial estimate and the signal-to-noise-ratio (SNR) decreases.
When the channel gets better, then the potential is wasted.
Several methods have been developed to combat these errors.
Their applications might need to be reevaluated and adjusted.

Modern transmission schemes that utilize channel information between several antennas and users suffer severely from movement \cite{Artner2013}.
Their performance might increase drastically with channel static antennas.

Channel static antennas allow a separation of wireless communication channels by their cause, i.e. changes that are caused by platform movement and channel changes that are caused by the environment, e.g. for vehicular channels \cite{Mecklenbrauker2011ProcIEEE}.

I am confident that antennas can be generally used to keep wireless communication channels static, and not only as a compensation for platform movement.
Patent pending \cite{Artner2018Patentanmeldung}.

\section*{Acknowledgment}


\noindent
The author thanks C.F.~Mecklenbr\"auker of Technische Universit\"at Wien, Vienna, Austria for fruitful discussions on applications.
The author thanks R.~Langwieser and S.~Pratschner, both of Technische Universit\"at Wien, Vienna, Austria, for their help with the experimental work.

\ifCLASSOPTIONcaptionsoff
  \newpage
\fi

\begin{IEEEbiography}[{\includegraphics[width=1in,height=1.25in,clip,keepaspectratio]{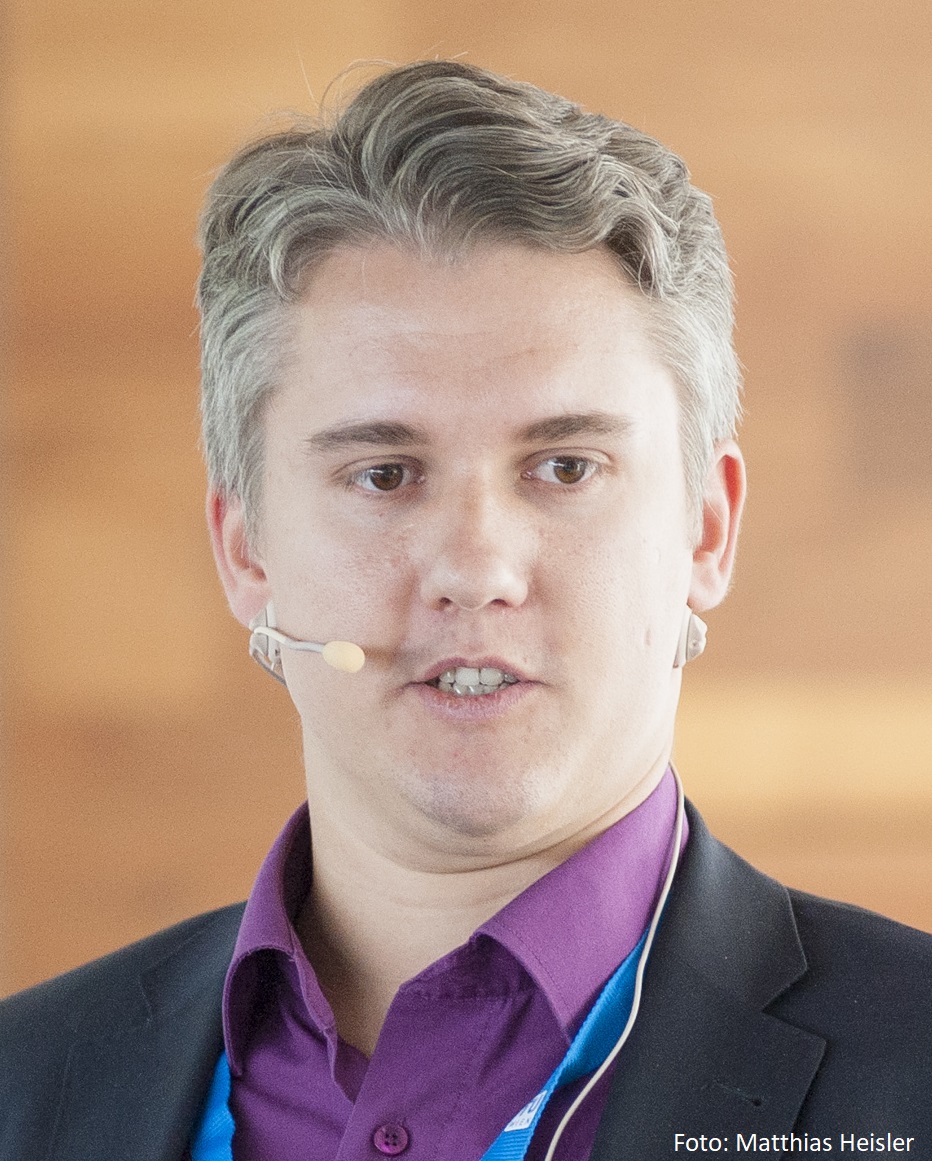}}]{Gerald Artner}
(M’13) was born in St.~P\"olten, Austria, in 1987.
He received the degree in electronic engineering with a specialization in computer engineering from the secondary technical school in St.~P\"olten (HTL), in 2007, the B.Sc. degree in electrical engineering and information technology, the Dipl.Ing. (M.Sc.) degree in telecommunications, and the Dr.techn. (Ph.D.) degree in electrical engineering from the Technische Universit\"at Wien, Vienna, Austria, in 2012, 2013, and 2017, respectively.
He is currently a university assistant with the Institute of Telecommunications, Technische Universit\"at Wien. His research interests include interference alignment, wireless communication testbeds, vehicular communications, automotive antennas, and carbon fiber reinforced polymer in antenna applications.
\end{IEEEbiography}

\end{document}